\documentclass[twocolumn,showpacs,preprintnumbers,amsmath,amssymb,floatfix,prb,showkeys]{revtex4}
\usepackage{graphicx}% Include figure files
\usepackage{epsfig}
\usepackage{dcolumn}% Align table columns on decimal point
\usepackage{bm}% bold math
\begin{document}

\preprint{}

\title{Spin Qubits in Multi-Electron Quantum Dots}

\author{Serguei Vorojtsov,$^1$ Eduardo R. Mucciolo,$^{1,2}$ and
Harold U. Baranger$^1$}

\affiliation{$^1$Department of Physics, Duke University, Box 90305, Durham, North
Carolina 27708-0305\\ $^2$Departamento de F\a'{\i}sica, Pontif\a'{\i}cia
Universidade Cat\'olica do Rio de Janeiro, C.P. 37801, 22452-970 Rio
de Janeiro, Brazil}

\date{\today}

\begin{abstract}
We study the effect of mesoscopic fluctuations on the magnitude of
errors that can occur in exchange operations on quantum dot
spin-qubits. Mid-size double quantum dots, with an odd number of
electrons in the range of a few tens in each dot, are investigated
through the constant interaction model using realistic parameters. It
is found that the constraint of having short pulses and small errors
implies keeping accurate control, at the few percent level, of several
electrode voltages. In practice, the number of independent parameters
per dot that one should tune depends on the configuration and ranges
from one to four.
\end{abstract}


% 03.67.Lx   Quantum computation
% 85.35.Gv   Single electron devices
% 73.21.La   Quantum dots


\maketitle

\onecolumngrid
\vspace*{-0.12in}
\twocolumngrid

%%%%%%%%%%%%%%%%%%%%%%%%%%%%%%%%%%%%%%%%%%%%%%%%%%%%%%%%%%%%%%%%%%%%%%%%%%%%%%%
\section{Introduction}

Since the discovery that quantum algorithms can solve certain
computational problems much more efficiently than classical
ones,\cite{shor94,grover96} attention has been devoted to the physical
implementation of quantum computation (QC). Among the many proposals,
there are those based on the spin of electrons in laterally confined
quantum dots (QD),\cite{loss98} which may have great potential for
scalability and integration with current technologies. For any
successful proposal, one must be able to perform single- and
double-qubit operations much faster than the decoherence time. In
fact, all logical operations required for QC can be realized if these
elementary operations are sufficiently error free.\cite{barenco95}

Single qubit operations involving a single QD will likely require
precise engineering of the underlying material or control over local
magnetic fields;\cite{divincenzo99} both have yet to be achieved in
practice. Two-qubit operations, in contrast, are already within
experimental reach. They can be performed by sending electrical pulses
to modulate the potential barrier between adjacent QDs. That permits
direct control over the effective, Heisenberg-like, exchange
interaction between the qubit spins, which is created by the overlap
between the electronic wave-functions of the QDs.\cite{loss98} These
operations are important elements in forming a basic two-qubit gate
such as the controlled-not \cite{divincenzo95} and in the propagation
of quantum information through QD arrays.\cite{loss98} In fact, using
three QDs instead of just one to form a logical qubit would allow one
to perform all logical operations entirely based on the exchange
interaction.\cite{divincenzo00} Thus, exchange operations will likely
play a major role in the realization of QD qubits. A quantitative
understanding of errors that occur during an exchange operation will
help in designing optimal systems.

The first proposal for a QD spin qubit\cite{loss98} relied on having a
single electron in a very small laterally confined QD. One advantage
of such a system is that the Hilbert space is nominally
two-dimensional. Leakage from the computational space involves
energies of order either the charging energy or the single-particle
excitation energy, both of which are quite large in practice ($\sim
1$~meV $\sim 10$~K). Working adiabatically -- such that the inverse of
the switching time is much less than the excitation energy -- assures
minimal leakage. The large excitation energy implies that pulses of
tens of picoseconds would be both well within the adiabatic regime and
below the dephasing time $\tau_\phi$ (which is typically in the
nanosecond range since orbital degrees of freedom are
involved). However, in practice, it is difficult to fabricate very
small tunable devices.\cite{Marcus_private} Moreover, one-electron QDs
may offer little possibility of gate tuning due to their rather
featureless wave functions.

Alternatively, a qubit could be formed by the top most ``valence''
electron in a QD with an odd number of electrons.\cite{hu01a} In this
case, electrons filling the lower energy states should comprise an
inert shell, leaving as the only relevant degree of freedom the spin
orientation of the valence electron. Large QDs with 100-1000
electrons, while much simpler to fabricate than single electron QDs,
are unsuitable because the excitation energy is small ($\sim
50\,\mu$eV $\sim 0.6$~K), leading to leakage or excessively slow
exchange operations. On the other hand, mid-size QDs, with 10-40
electrons, are sufficiently small to have substantial excitation
energies, yet both reasonable to fabricate and tunable through plunger
electrodes. For these dots, a careful analysis of errors is necessary.

Perhaps the best example of an exchange operation is the swap of the
spin states of the two qubits. For instance, it causes up-down spins
to evolve to down-up. Maximum entanglement between qubits occurs when
half of a swap pulse takes place -- a square-root-of-swap operation.
Several authors have treated the problem of swap errors in QD
systems.\cite{burckard99,hu00-02,hu01,schliemann01,brandes02} A
primary concern was the occurrence of double occupancy (when both
valence electrons move into the same QD) during and after the
swap. However, no study so far has considered another intrinsic
characteristic of electronic states in multi-electron QDs, namely,
their marked dependence on external perturbations such as electrode
voltage or magnetic field. This sensitivity gives rise to strong
sample-to-sample fluctuations arising from the phase-coherent orbital
motion.\cite{mesoscopics} These features can make the precise control
of energy levels, wave functions, and inter-dot couplings a difficult
task.

In this work we study errors and error rates that can take place
during the exchange operation of two spin qubits based in
multi-electron QDs. We consider realistic situations by taking into
account an extra orbital level and fluctuations in level positions and
coupling matrix elements. These lead to deviations from a
pre-established optimal swap operation point, especially when a
single-particle level falls too close to the valence electron
level. Reasons for such fluctuations can be, for instance, (i) the
lack of a sufficient number of tuning parameters (i.e., plunger
electrodes), or (ii) the cross-talk between the tuning electrodes. Our
results set bounds on the amount of acceptable detuning for mid-size
QD qubits.

This paper is organized as follows. In Sec. \ref{sec:model}, we
introduce and justify the model Hamiltonian. The states involved in
the exchange operation are presented in Sec. \ref{sec:error}, where we
also discuss the pulses and the parameters involved in the exchange
operations. In Sec. \ref{sec:results} we present the results of our
numerical simulations. We also discuss the impact of mesoscopic
effects on errors and put our analysis in the context of actual
experiments. Finally, in Sec. \ref{sec:conclusion} we draw our
conclusions.

%%%%%%%%%%%%%%%%%%%%%%%%%%%%%%%%%%%%%%%%%%%%%%%%%%%%%%%%%%%%%%%%%%%%%%%%%%%%%%%
\section{Model System}
\label{sec:model}

We begin by assuming that the double QD system can be described by the
Hamiltonian\cite{orthodox}
\begin{equation}
\label{eq:hamilton}
H = H_A + H_B + H_{AB},
\end{equation}
where
\begin{equation}
\label{eq:CImodel}
H_\alpha = \sum_{j,\sigma} \epsilon_j^\alpha\, n_{\alpha,j\sigma} +
\frac{U_\alpha}{2} \sum_{j,\sigma} n_{\alpha,j\sigma} \Big(
\sum_{k,\sigma^\prime} n_{\alpha,k\sigma^\prime} - 1 \Big),
\end{equation}
$\alpha = A,B$, and
\begin{equation}
\label{eq:hopping}
H_{AB} = \sum_{j,k,\sigma} \left( t_{jk}\, a_{j\sigma}^\dagger
b_{k\sigma} + \text{h.c.} \right).
\end{equation}
Here, $n_{A,j\sigma} = a^\dagger _{j\sigma} a_{j\sigma}$ and
$n_{B,k\sigma} = b^\dagger _{k\sigma} b_{k\sigma}$ are the number
operators for the single-particle states in the QDs (named $A$ and
$B$), $\epsilon^\alpha_j$ denotes the single-particle energy levels,
$t_{jk}$ are the tunneling amplitudes between the dots, and
$U_{\alpha}$ is the charging energy ($\sigma = \uparrow, \downarrow$
and $j,\,k$ run over the single-particle states). Typically, for mid-
to large-size QDs, the charging energy is larger than the mean level
spacing.

In the literature of Coulomb blockade phenomena in closed QDs, the
Hamiltonian in Eq.~(\ref{eq:CImodel}) is known as the constant
interaction model. It provides an excellent description of
many-electron QDs, being supported by both microscopic calculations
and experimental data.\cite{mesoscopics, orthodox, ullmo01a, usaj,
ullmo01b}
The reasoning behind its success can be understood from two
observations. First, mid- to large-size QDs, with more than ten
electrons, behave very much like conventional disordered metals in the
diffusive regime. Wavelengths are sufficiently small to resolve
irregularities in the confining and background potentials, leading to
classical chaos and the absence of shell effects in the energy
spectrum. In this case, the single-particle states obey the statistics
of random matrices, showing complex interference patterns and
resembling a random superposition of plane waves. This is in contrast
with the case of small, circularly symmetric, few-electron QDs, where
shell effects are pronounced.\cite{hu01a,kouwenhoven01}

Second, for realistic electron densities, the QD linear size is larger
than the screening length of the Coulomb interactions. In the presence
of random plane waves, the screened interaction can then be broken up
into a leading electrostatic contribution characterized by the QD
capacitance plus weak inter-particle residual
interactions.\cite{orthodox,ullmo01b} This description becomes more
accurate as the number of electrons gets larger since the residual
interactions become weaker. The electron bunching is reduced as wave
functions become more uniformly extended over the QD. Also, the
increase in the number of oscillations in the wave functions leads to
a self-averaging of the residual interactions. In this limit, one
arrives at the so-called ``universal Hamiltonian" for QDs, containing
only single-particle levels, the charging energy, and a mean-field
exchange term.\cite{mesoscopics,orthodox} This Hamiltonian can be
derived explicitly via a random-phase approximation treatment of the
Coulomb interaction and the use of random-matrix wave
functions.\cite{orthodox,ullmo01b}

According to these arguments, interaction effects beyond the charging
energy term are omitted in Eq. (\ref{eq:CImodel}). In addition, the
intra-dot exchange interaction, which tends to spin polarize the QD,
is also neglected. The reason for that is the following. One can show
that the intra-dot exchange term only affect states where there is
double occupancy of a level. Thus, the exchange interaction constant
always appears side-by-side with the charging energy. But in
multi-electron dots, the exchange energy (which is at most of order of
the mean level separation) is much smaller than the charging
energy. Thus, intra-dot exchange effects are strongly suppressed by
the charging energy. We have verified that their inclusion does not
modify appreciably our final results. We expect the exchange
interaction to become important for two-qubit operations only in the
case of small QDs with only a few electrons, when all energy scales
(including the mean level spacing) are of the same order.

Thus, the simple picture where single-particle states are filled
according to the Pauli principle up to the top most level is an
appropriate description of multi-electron dots.\cite{mesoscopics,
orthodox, ullmo01a, usaj, ullmo01b} In order to define the spin-$\frac{1}{2}$
qubits, both QDs should contain an odd number of electrons (say,
$2N_{A}-$1 and $2N_{B}-$1). The QD spin properties are then dictated
by the lone, valence electron occupying the highest level. The
remaining electrons form an inert core, provided that operations are
kept sufficiently slow so as not to cause particle-hole excitations to
other levels.

Experimentally, the two-qubit exchange operations also require the
capability of isolating the QDs from each other, so that a direct
product state can be prepared, such as
\begin{equation}
|i\rangle = |N_A,\uparrow\rangle_A\otimes |N_B,\downarrow\rangle_B,
\end{equation}
where the kets represent only the spin of the valence electron on each
QD.

%%%%%%%%%%%%%%%%%%%%%%%%%%%%%%%%%%%%%%%%%%%%%%%%%%%%%%%%%%%%%%%%%%%%%%%%%%%%%%%
\section{Errors in Exchange Operations}
\label{sec:error}

We focus our study on errors that appear after a full swap operation
(which should result in no entanglement). Although it could in
principle seem more sensible to look at the square-root-of-swap
operation (which creates entanglement and is therefore a building
block of logical gates), error magnitudes for the latter are
straightforwardly related to those of the full swap operation, as we
will show. We leave the discussion of the square root of swap to
Sec. \ref{sec:results}.

The ideal full swap operation exchanges the valence electrons of the
QD system. For instance, it takes the product state $|i\rangle$ into
\begin{equation}
|f\rangle = \hat{U}_\text{sw}\,
|i\rangle = |N_A,\downarrow\rangle_A\otimes
|N_B,\uparrow\rangle_B.
\end{equation}
Physically, the full swap can be implemented by starting with isolated
QDs, turning on the inter-dot coupling for a time $T$ (the pulse
duration), and then turning it off, isolating the QDs again. For
weakly coupled QDs ($|t| \ll U,\delta\epsilon$), one finds $T \approx
(\pi/4)\, U/|t|^{2}$, where $t$ and $U$ here represent typical values
for the coupling matrix element and the charging energy, respectively
(throughout we assume $\hbar = 1$). To quantify the amount of error
that takes place during the operation, we use the probability of not
reaching $|f \rangle$ asymptotically, namely,
\begin{equation}
\label{eq:error}
\varepsilon = 1 - | \langle f | \psi ( + \infty ) \rangle|^2.
\end{equation}
%

%%%%%%%%%%%%%%%%%%
\begin{figure}
\includegraphics[width=5.0cm]{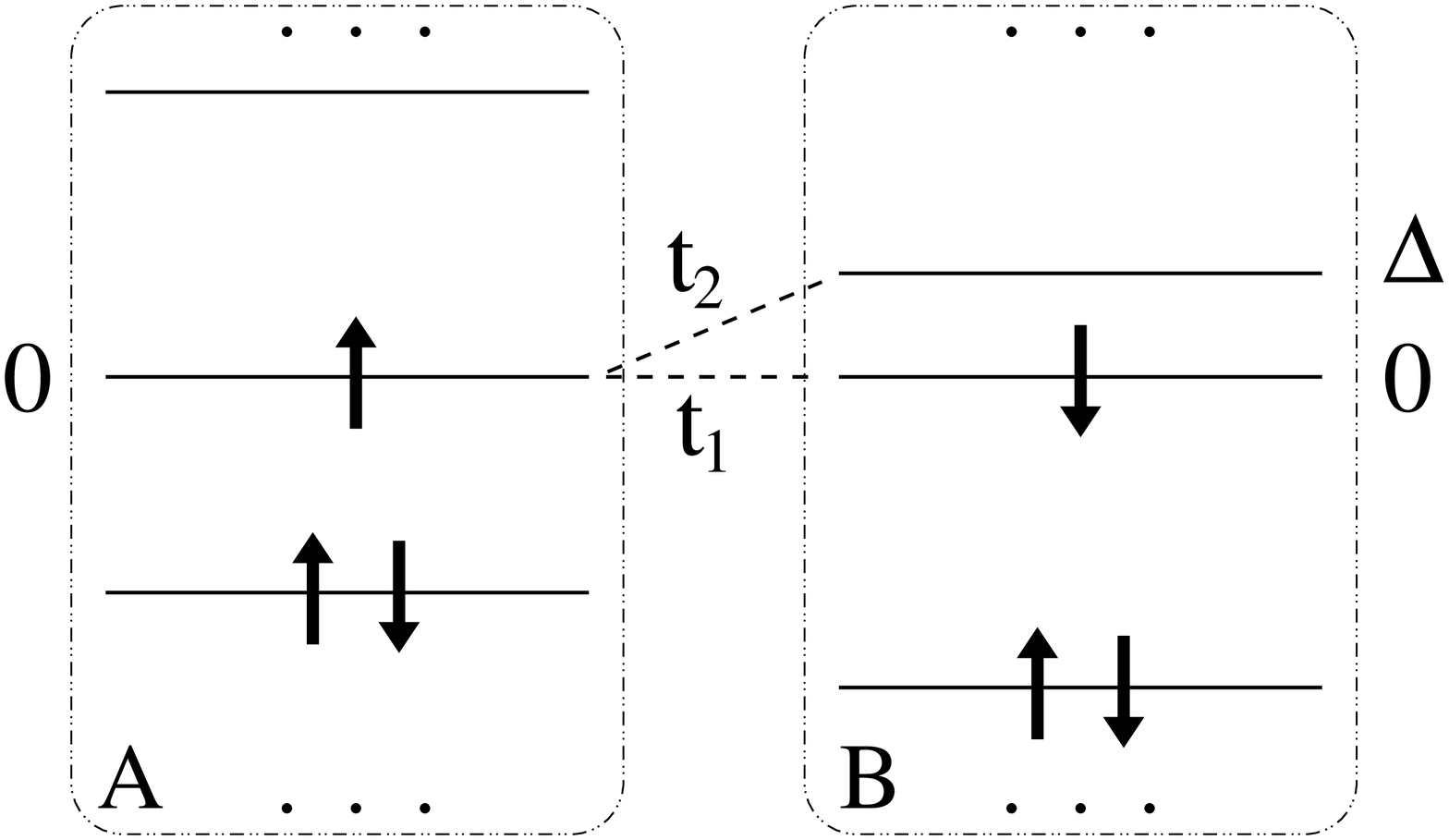}
\caption{Schematic disposition of energy levels of a system of 
two QD spin qubits (only levels close to the top occupied state 
are shown). The dashed lines indicate the most probable transitions 
that can occur during the exchange operation.}
\label{fig:1}
\end{figure}  
%%%%%%%%%%%%%%%%%%

We solved numerically the time-dependent Schr\"odinger equation that
derives from Eq. (\ref{eq:hamilton}) for the particular but
nevertheless realistic case shown in Fig. \ref{fig:1}. We assumed that
voltage tuning allows one to place the top most electron of QD $A$
into an isolated single-particle state of energy $\epsilon^{A}_{N_A} =
0$ aligned with the energy of the top most electron in QD $B$,
$\epsilon^{B}_{N_B} = 0$. However, limited tuning ability leaves an
adjacent empty state close in energy in QD $B$: $\epsilon^{B}_{N_B+1}
= \Delta$. Therefore, while we can approximately neglect all levels
but one in QD $A$, for QD $B$ we needed to take two levels into
account, having hopping matrix elements denoted by $t_1=t_{N_{A}N_B}$
and $t_2=t_{N_A,N_B+1}$.\cite{timers} To facilitate the analysis, we
assumed that the dots have the same capacitance, $C$, so that
$U_A=U_B=U=e^2/C$.

The Hamiltonian of Eq. (\ref{eq:hamilton}) conserves total
spin. Assuming that filled inner levels in both QDs are inert (forming
the ``vacuum'' state $| 0 \rangle$), we can span the $S_z=0$ Hilbert
subspace with nine two-electron basis states. According to their
transformation properties, they can be divided into ``singlet''
\begin{eqnarray}
|S_l \rangle & = & \frac{1}{\sqrt{2}} \left(
b_{N_B+l-1,\downarrow}^\dagger a_{N_A\uparrow}^\dagger -
b_{N_B+l-1,\uparrow}^\dagger a_{N_A\downarrow}^\dagger \right) |0
\rangle, \nonumber \\ |D_l \rangle & = &
b_{N_B+l-1,\downarrow}^\dagger b_{N_B+l-1,\uparrow}^\dagger | 0
\rangle, \\ |D_3 \rangle & = & \frac{1}{\sqrt{2}} \left(
b_{N_B+1,\downarrow}^\dagger b_{N_B\uparrow}^\dagger -
b_{N_B+1,\uparrow}^\dagger b_{N_B\downarrow}^\dagger \right) |
0\rangle, \nonumber \\ |D_4 \rangle & = & a_{N_A\downarrow}^\dagger
a_{N_A\uparrow}^\dagger | 0 \rangle, \nonumber
\end{eqnarray}
and ``triplet''
\begin{eqnarray}
| T_l \rangle & = & \frac{1}{\sqrt{2}} \left(
b_{N_B+l-1,\downarrow}^\dagger a_{N_A\uparrow}^\dagger +
b_{N_B+l-1,\uparrow}^\dagger a_{N_A\downarrow}^\dagger \right) | 0
\rangle, \nonumber \\ |D_5 \rangle & = & \frac{1}{\sqrt{2}} \left(
b_{N_B+1,\downarrow}^\dagger b_{N_B\uparrow}^\dagger +
b_{N_B+1,\uparrow}^\dagger b_{N_B\downarrow}^\dagger \right) |
0\rangle,
\end{eqnarray}
classes, with $l=1,2$. 

The final states that correspond to an error have either double
occupancy ($|D_k\rangle$, $k=1,\ldots,5$), or an electron in the
$(N_B+1)$-level of QD $B$ ($|S_2\rangle$ and
$|T_2\rangle$).\cite{obs2} In addition, a return to the initial state
is also considered an error. It is worth noticing the difference
between our treatment of the problem and that of
Ref. \onlinecite{schliemann01}. In our case, errors come mainly from
either ending in the excited single-particle state after the operation
is over (i.e., states $|S_2\rangle$ and $|T_2\rangle$), or from
``no-go" defective operations. In Ref. \onlinecite{schliemann01},
errors come from having double occupancy in the final state. Double
occupancy errors can be exponentially suppressed by adiabatically
switching the pulse on and off on time scales larger than the inverse
charging energy.\cite{loss98,hu00-02,schliemann01} Making pulses
adiabatic on the time scale of the inverse mean level spacing for
multi-electron quantum dots is more challenging, especially because
the spacings fluctuate strongly both from quantum dot to quantum dot
and upon variation of any external parameter (mesoscopic
fluctuations). Therefore, multi-electron quantum dots require extra
tunability to get around such problems.

Very small errors, below $10^{-6}$--$10^{-4}$, can, in principal, be
fixed by the use of error correction algorithms.\cite{steane98} The
pulses, therefore, should be sufficiently adiabatic for errors to
remain below this threshold. We adopted the following pulse shape:
\begin{equation}
\label{eq:pulse}
v(t) = \frac{1}{2} \left( \tanh\frac{t+T/2}{2\tau}-
\tanh\frac{t-T/2}{2\tau} \right),
\end{equation}
where $\tau$ is the switching time. The pulse will remain both
well-defined and adiabatic provided that $T \gg \tau \gg \text{max} \{
\Delta^{-1}, U^{-1} \}$. Notice that this pulse is equivalent to that
adopted in Ref. \onlinecite{schliemann01} up to exponential accuracy,
$O(e^{-T/\tau})$, with $T\gg\tau$. There is no particular reason to
believe that either performs better than the other; our choice was
dictated by technical convenience.

%%%%%%%%%%%%%%%%%%
\begin{figure}
\includegraphics[width=7.4cm]{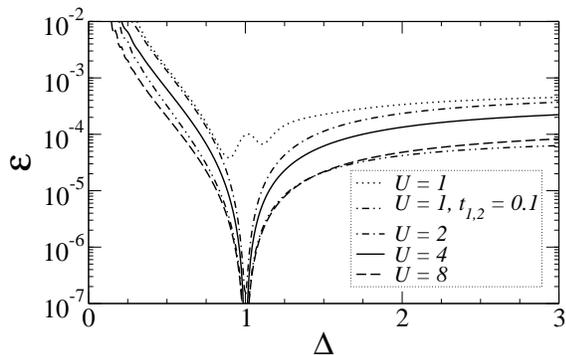}
\caption{Full swap error as a function of upper level detuning in
quantum dot $B$. The pulse width is optimized for $\Delta = 1$, $\tau
= 6$, and $t_{1,2} = 0.2$. Results for different charging energies are
shown. Interference between different quantum mechanical paths in the
device causes a sharp minimum.}
\label{fig:2}
\end{figure}  
%%%%%%%%%%%%%%%%%%

%%%%%%%%%%%%%%%%%%%%%%%%%%%%%%%%%%%%%%%%%%%%%%%%%%%%%%%%%%%%%%%%%%%%%%%%%%%%%%%
\section{Results}
\label{sec:results}

We used a standard numerical method, the so-called Richardson
extrapolation,\cite{press92} to solve the Schr\"odinger equation for
$|\psi(t)\rangle$. The first step in our analysis was to find the
optimal value of $T$ which minimized the full swap error, as defined
in Eq. (\ref{eq:error}), for a given set of parameters $U$, $t_1$, and
$\tau$ (we used $\Delta=1$ and took $t_2 = t_1$). The second step was
to study how this minimal error depends on $\tau$. There is actually
an optimal interval for $\tau$, since small switching times spoil
adiabaticity, while large ones compromise the pulse shape (when $T$ is
relatively short). Empirically, we find that errors related to
switching times become negligible once $\tau$ reaches values of about
$\tau_0 = 4\, \text{max} \{ \Delta^{-1}, U^{-1} \}$, provided that
$\tau \ll T$. In what follows, we fix $\tau \ge \tau_0$.

%%%%%%%%%%%%%%%%%%
%\begin{figure}
%\includegraphics[width=8.5cm]{fig2plus3.eps}
%\caption{\footnotesize {Full swap error as a function of: (a) upper
%level detuning in quantum dot $B$; (b) detuning in the coupling
%constant $t_2$. The pulse width is optimized for $\Delta = 1$, $\tau = 6$, 
%and $t_{1,2} = 0.2$. Results for different charging energies are shown.}}
%\label{fig:2}
%\end{figure}  
%%%%%%%%%%%%%%%%%%

%%%%%%%%%%%%%%%%%%%%%%%%%%%%%%%%%%%%%%%%%%%%%%%%%%%%%%%%%%%%%%%%%%%%%%%%%%%%%%%
\subsection{Mesoscopic Effects}

Figure \ref{fig:2} shows the full swap error as a function of $\Delta$
when $T$ is fixed to its optimal value for $\Delta = 1$. Such a
situation would arise experimentally if the pulse is optimized for a
certain configuration, but a fluctuation in level spacing
occurs. Notice the sharp increase in error as $\Delta$
decreases. While increasing $\tau$ reduces this error (by making the
switching more adiabatic), very small level spacings would be
problematic, since $\tau$ can not be larger than $T$ without
sacrificing pulse shape and effectiveness. In order to make space for
an adiabatic switching time for small $\Delta$, one would also have to
increase pulse duration. This is clear in the case of $U = 1$ (see
Fig. \ref{fig:2}): Even moderate couplings, $t_{1,2} = 0.2$, lead to
larger errors, which can then be suppressed by decreasing $t_{1,2}$ by
a factor of two; however, that causes a fourfold increase in pulse
width which may be problematic in terms of decoherence.

\begin{figure}
\includegraphics[width=7.4cm]{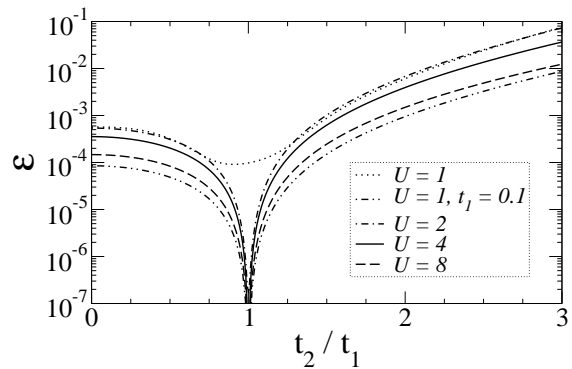}
\caption{Full swap error as a function of detuning in the coupling
constant $t_2$. The parameters used in the pulse width optimization
are the same as in Fig.~\ref{fig:2}. Results for different charging
energies are shown.}
\label{fig:3}
\end{figure}

The dependence of errors on fluctuations in the coupling amplitude
$t_2$ is shown in Fig.~\ref{fig:3}. Again, the pulse duration used is
the optimal value obtained when $t_2 = t_1 = 0.2$. As expected, the
error grows as $t_2$ increases. Errors related to large values of
$t_2$ can also be minimized by increasing the switching time, but the
same issues raised above appear.

%%%%%%%%%%%%%%%%%%
%%%%%%%%%%%%%%%%%%

\begin{figure}
\includegraphics[width=7.4cm]{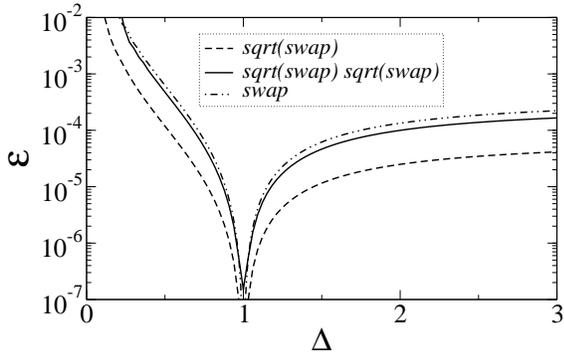}
\caption{Comparison of error resulting from a full swap operation and
two consecutive square root of swap operations. The error is plotted
as a function of upper level detuning in quantum dot $B$. The pulse
width is optimized for $\Delta =1$, $t_{1,2}=0.2$, $U=4$, and $\tau
=6$. The error for a single square-root-of-swap operation is also
shown.}
\label{fig:6}
\end{figure}

Figure~\ref{fig:6} presents the error for two situations involving
pulses with duration of about $T/2$, corresponding to the
square-root-of-swap operation. The cases shown are: (i) one, and (ii)
two consecutive square-root-of-swap pulses. For comparison, the curve
corresponding to a full swap pulse is also shown. The error for the
square-root-of-swap operation is given by Eq. (\ref{eq:error}) with
$|f\rangle$ replaced by
\begin{equation}
\label{eq:fsqrt}
\left| f^\prime \right> = \frac{1-i}{2}\left| S_{1} \right>
+ \frac{1+i}{2}\left| T_{1} \right> .
\end{equation}
One can observe from Fig. \ref{fig:6} that error rates are nearly the
same after a full swap operation and after two consecutive square root
of swap operations. This insensitivity of the error to the pulse
duration led us to concentrate our effort on the full swap operations.

%%%%%%%%%%%%%%%%%%
%%%%%%%%%%%%%%%%%%

In order to establish an upper bound for QD tuning accuracy, we have
performed simulations where both $\Delta$ and $t_2$ were allowed to
vary. The spacing between the levels in QD $B$ was taken from a
Gaussian distribution centered at $\bar{\Delta}=1$, with standard
deviation $\sigma_\Delta$ ($\epsilon_{N_B}^{B}=0$ was kept fixed). For
the coupling amplitude, we generated Gaussian distributed level widths
$\Gamma_2 = 2\pi\, t_2^2/\bar{\Delta}$, with average $\bar{\Gamma}_2 =
2\pi\, t_1^2/\bar{\Delta}$ and standard deviation $\sigma_{\Gamma_2}$.
The pulses had their widths optimized for the typical case where
$\Delta = \bar{\Delta} = 1$, $U = 4$, $t_{1,2} = 0.2$, and $\tau =
6$. For fixed values of $\bar{\Delta}$, $\sigma_\Delta$,
$\bar{\Gamma}_2$, and $\sigma_{\Gamma_2}$, we generated 10,000
realizations of $\Delta$ and $\Gamma_2$ and each time calculated the
error after the application of the full swap pulse. In
Fig.~\ref{fig:5} we show how the probability of having an error larger
than the $10^{-4}$ threshold depends on the energy level accuracy,
$\sigma_\Delta$. Two cases are considered, namely, plain and limited
control of the inter-dot coupling constant ($\sigma_{\Gamma_2} = 0$
and $0.1\, \bar{\Gamma}_2$, respectively). The data indicates that
frequent, non-correctable errors will happen if an accuracy in
$\Delta$ of better than 10 percent is not achieved.

%%%%%%%%%%%%%%%%%%%%%%%%%%%%%%%%%%%%%%%%%%%%%%%%%%%%%%%%%%%%%%%%%%%%%%%%%%%%%%%
\subsection{Relevance for Real Quantum Dots}

\begin{figure}
\includegraphics[width=7.4cm]{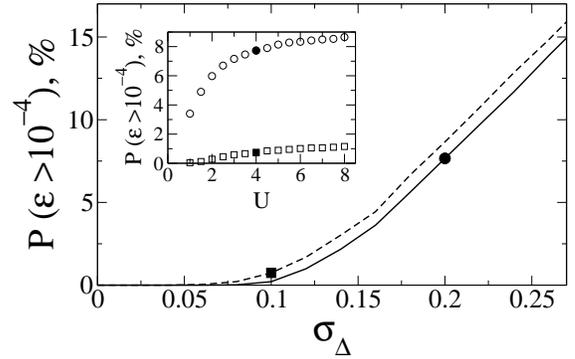}
\caption{Probability of having excessively large full swap errors 
(percentage) as a function of level spacing detuning. The solid (dashed) 
line corresponds to $\sigma_{\Gamma_2}/\bar{\Gamma}_2 = 0$ ($0.1$) 
at $U=4$ and $t_1=0.2$. The inset shows how the probability varies for 
a fixed width pulse ($T \approx 90$), but different charging energies $U$, 
when there is a large level-position detuning:
$\sigma_\Delta = 0.2$, $\sigma_{\Gamma_2} = 0$ (circles) and moderate
level-position and coupling detunings: $\sigma_\Delta = 0.1$,
$\sigma_{\Gamma_2}/\bar{\Gamma}_2 =0.1$ (squares).}
\label{fig:5}
\end{figure}

To make a quantitative estimate of the impact of these results, let us
consider the double QD setup of Jeong and coworkers.\cite{jeong01} In
their device, each QD holds about 40 electrons and has a lithographic
diameter of 180~nm (we estimate the effective diameter to be around
120~nm, based on the device electron density). The charging energy and
mean level spacing of each QD are approximately 1.8~meV and 0.4~meV,
respectively (thus $U/\bar{\Delta} \approx 4.5$). If we allow for a
maximal inter-dot coupling of $t_{1,2} \approx 0.2\, \bar{\Delta}$
(which yields a level broadening of about $0.25\, \bar{\Delta}$), we
find minimal full swap pulse widths of about 100~ps. These values
match those used in Fig.~\ref{fig:5}. For this case, switching times
of 10~ps would be long enough to operate in the adiabatic regime and
also provide an efficient and well-defined pulse shape. Thus, the
combined times should allow for 8-10 consecutive full swap gates
before running into dephasing effects related to orbital degrees of
freedom.\cite{fujisawa03} While these numbers are yet too small for
large-scale quantum computation, they could be sufficient for the
demonstration of QD spin qubits. Based on Fig.~\ref{fig:5}, we find
that accuracies in $\Gamma_2$ of about 10\% would make operations only
limited by dephasing, and not by fluctuation-induced errors. However,
as shown in the inset, even for small QDs (typically having small
$U/\bar{\Delta}$ ratios), the occurrence of large errors is quite
frequent when level detuning is large.

%%%%%%%%%%%%%%%%%%
%%%%%%%%%%%%%%%%%%

An important issue for multi-electron QDs is their strong mesoscopic,
sample-to-sample fluctuations in energy level position and
wave-function amplitudes. Our results so far indicate how big an
effect a given change in energy or wave-function will produce; now we
go further and discuss how mesoscopic fluctuations more generally
affect a collection of qubits.

In experiments, several electrodes are placed around the QD
surroundings and their voltages are used to adjust the lateral
confining potential, the inter-dot coupling, and the coupling between
QDs and leads. These voltages are external parameters that can be used
to mitigate the effects of mesoscopic fluctuations by tuning energy
levels and wave-functions to desired values. Having that in mind, our
results indicate two different scenarios for QD qubit implementations.

First, if one is willing to characterize each QD pair separately and
have them operate one by one, mesoscopic fluctuations will be
irrelevant. It will be possible, with a single parameter per QD, say,
to isolate and align energy levels reasonably well. Errors can be
further minimized by decreasing the inter-dot coupling (thus
increasing $T$). But since QDs are not microscopically identical, each
pair of QDs will require a different pulse shape and
duration. Multi-electron QDs are tunable enough, easy to couple, and
much easier to fabricate than one-electron dots; therefore,
multi-electron QDs are most appropriate for this case.

Second, if the goal is to achieve genuine scalability, one has to
operate qubits in a similar and uniform way, utilizing a single pulse
source. In this case, $T$ and $\tau$ should be the same for all QD
pairs. Based on our results above, one should strive to maximally
separate the top most occupied state from all other states, occupied
or empty, so as to reduce the possibility of leakage during operations
with a fixed duration. At the same time, it is important to reduce
inter-pair cross-talk induced by capacitive coupling between
electrodes, as well as all inter-dot couplings except between the top
most states of each QD. One should bear in mind that not all
electrodes act independently -- in most cases a search in a
multidimensional parameter space has to be carried out. Thus, four
tuning parameters per QD may be necessary to achieve the following
goals: (i) find isolated, single-occupied energy level (two
parameters); (ii) align this level with the corresponding level in an
adjacent QD (one parameter); (iii) control the inter-dot coupling (one
parameter). For parameters involved in (i) and (ii), an accuracy of a
few percent will likely be required. Finally, control over the
inter-dot coupling parameter, (iii), must allow for the application of
smooth pulse shapes in the picosecond range. Our simulations also show
that the pulse width must be controlled within at least $0.5\%$
accuracy. Although these requirements seem quite stringent, recent
experiments indicate that they could be met.\cite{chen03}

%%%%%%%%%%%%%%%%%%%%%%%%%%%%%%%%%%%%%%%%%%%%%%%%%%%%%%%%%%%%%%%%%%%%%%%%%%%%%%%
\section{Conclusions}
\label{sec:conclusion}

In summary, our analysis indicate that mid-size QDs, with ten to a few
tens of electrons, while not allowing for extremely fast gates, are
still good candidates for spin-qubits. They offer the advantage of
being simpler to fabricate and manipulate, but at the same time
require accurate, simultaneous control of several parameters. Errors
related to detuning and sample-to-sample fluctuations can be large,
but can be kept a secondary concern with respect to dephasing effects
provided that a sufficient number of independent electrodes or tuning
parameters exists.

%%%%%%%%%%%%%%%%%%%%%%%%%%%%%%%%%%%%%%%%%%%%%%%%%%%%%%%%%%%%%%%%%%%%%%%%%%%%%%%
\begin{acknowledgments}
We thank D. P. DiVincenzo, C. M. Marcus, and G. Usaj for useful
discussions. This work was supported in part by the National Security
Agency and the Advanced Research and Development Activity under ARO
contract DAAD19-02-1-0079. Partial support in Brazil was provided by
Instituto do Mil\^enio de Nanoci\^encia, CNPq, FAPERJ, and PRONEX.
\end{acknowledgments}

%%%%%%%%%%%%%%%%%%%%%%%%%%%%%%%%%%%%%%%%%%%%%%%%%%%%%%%%%%%%%%%%%%%%%%%%%%%%%%
\onecolumngrid
\vspace*{-0.15in}
%\vfill
\twocolumngrid

%%%%%%%%%%%%%%%%%%%%%%%%%%%%%%%%%%%%%%%%%%%%%%%%%%%%%%%%%%%%%%%%%%%%%%%%%%%%%%%


\begin{thebibliography}{11}
\bibitem{timers} Though we assume time-reversal symmetry (real $t_{1}$
and $t_{2}$), we have checked that its violation does not modify our
results.
\bibitem{obs2} While such states can lead to the full spin swap (with
probability equal to $1/2$), they will induce severe detuning errors
in subsequent operations.
\end{thebibliography}
\end{document}